\documentclass[aps,pra,twocolumn,floatfix,groupedaddress,superscriptaddress,footinbib,notitlepage]{revtex4-2}
\usepackage{times,graphics,graphicx,bm,bbm,bbold,amssymb,amsmath,amsfonts,dsfont,cancel,color,xcolor}
\usepackage[normalem]{ulem}
\usepackage[colorlinks=true,citecolor=blue,linkcolor=blue,urlcolor=blue]{hyperref}
\newcommand{\ignore}[1]{}
\usepackage[normalem]{ulem}

\newcommand\cysout{\bgroup\markoverwith{\textcolor{cyan}{\rule[0.5ex]{2pt}{1.2pt}}}\ULon}

\def\dbar{{\,\,\mathchar'26\mkern-12mu d\hskip0.2mm}}
\DeclareFontFamily{OT1}{pzc}{}
\DeclareFontShape{OT1}{pzc}{m}{it}
              {<-> s * [1.25] pzcmi7t}{}
\DeclareMathAlphabet{\mathpzc}{OT1}{pzc}
                                 {m}{it}
\usepackage[T1]{fontenc} 
\usepackage{newtxtext,newtxmath}
\begin{document}

\title{Entropy-Based Formulation of Thermodynamics in Arbitrary Quantum Evolution}

\author{S. Alipour}
\affiliation{QTF Center of Excellence, Department of Applied Physics, Aalto University, P. O. Box 11000, FI-00076 Aalto, Espoo, Finland}

\author{A. T. Rezakhani}
\affiliation{Department of Physics, Sharif University of Technology, Tehran 14588, Iran}

\author{A. Chenu}
\affiliation{Department of Physics and Materials Science, University of Luxembourg, L-1511 Luxembourg, G.D. Luxembourg}
\affiliation{Donostia International Physics Center,  E-20018 San Sebasti\'an, Spain}
\affiliation{IKERBASQUE, Basque Foundation for Science, E-48013 Bilbao, Spain}
%\affiliation{Current address: Department of Physics and Materials Science, University of Luxembourg, L-1511 Luxembourg, G.D. Luxembourg}

\author{A. del Campo}
\affiliation{Department of Physics and Materials Science, University of Luxembourg, L-1511 Luxembourg, G.D. Luxembourg}
\affiliation{Donostia International Physics Center,  E-20018 San Sebasti\'an, Spain}
\affiliation{IKERBASQUE, Basque Foundation for Science, E-48013 Bilbao, Spain}
\affiliation{Department of Physics, University of Massachusetts, Boston, MA 02125, USA}

\author{T. Ala-Nissila}
\affiliation{QTF Center of Excellence, Department of Applied Physics, Aalto University, P. O. Box 11000, FI-00076 Aalto, Espoo, Finland}
\affiliation{Interdisciplinary Centre for Mathematical Modelling and Department of Mathematical Sciences, Loughborough University, Loughborough, Leicestershire LE11 3TU, UK}

%%%%%%%%%%%%%%%%%%%%%%%%%%%%%%%%%%%%%%%%%%%%%%%%%%%%%%%%%%%%%%%%
\begin{abstract}
Given the evolution of an arbitrary open quantum system, we formulate a general and unambiguous method to separate the internal energy change of the system into an entropy-related contribution and a part causing no entropy change, identified as heat and work, respectively. We also demonstrate that heat and work admit geometric and dynamical descriptions by developing a universal dynamical equation for the given trajectory of the system. The dissipative and coherent parts of this equation contribute exclusively to heat and work, where the specific role of a work contribution from a counterdiabatic drive is underlined. Next we define an expression for the irreversible entropy production of the system which does not have explicit dependence on the properties of the ambient environment; rather, it depends on a set of the system's observables excluding its Hamiltonian and is independent of internal energy change. We illustrate our results with three examples.
\end{abstract}

\maketitle
%%%%%%%%%%%%%%%%%%%%%%%%%%%%%%%%%%%%%%%%%%%%%%%%%%%%%%%%%%%%%%%%

%%%%%%%%%%%%%%%%%%%%%%%%%%%%%%%%%%%%%%%%%%%%%%%%%%%%%%%%%%%%%%%%
Quantum thermodynamics holds a central stage at the interface of quantum information science, statistical mechanics, and quantum technologies, and it has shed new light on the laws of thermodynamics in the quantum regime \cite{gemmer2009quantum, book:Qthermo, book:Deffner-Campbell, Kosloff13, Goold16, Kieu04, Brandao15, Masanes17, HosseinNejad-OlayaCastro, SciRep:Alipour-corr, Marcantoni-entrop-nonMarkov, Esposito09, Campisi11, Funo18, Bakhshinezhad_2019}. Apart from uncovering a plethora of new phenomena, quantum thermodynamics has also motivated efforts to engineer efficient quantum machines in the laboratory \cite{Rossnagel16,Deng18Sci,Maslennikov19,Serra18,Poschinger18}.
 
Yet, several fundamental issues remain to be clarified. Most notably, except for particular regimes (e.g., weak-coupling, Markovian dynamics with slow driving \cite{Alicki}), an unequivocal definition of heat and work in arbitrary open-quantum system dynamics has been elusive. One problem is that such thermodynamic variables are not observables described by Hermitian operators \cite{Talkner08}; rather, they are trajectory-dependent quantities \cite{Vilar-Rubi-WorkDefNotCorrect, Aspuru-Guzik-ThermoInconsistencies, Niedenzu-ConceptsOfWork, Sampaio-ConditionalWaveF, Elouard17}. Existing definitions mostly incorporate generators of the dynamics (besides the state) and dynamical master equations with \textit{coherent} and \textit{dissipative} parts \cite{Alicki,Jarzynski-MeanForce,Gelin-MeanForce,Campisi-MeanForce-ParitionFunction,Miller-Anders-MeanForce,Rivas-nonEquilibMeanForce,SciRep:Alipour-corr}. For instance, in the widely-used \textit{conventional} framework \cite{Spohn-EP,Alicki} the internal energy change due to the dissipative (coherent) part of the master equation has been called heat (work) change. An alternative strategy uses the potential of mean force that amounts to pre-averaging the total partition function over the environmental degrees of freedom \cite{Jarzynski-MeanForce, Gelin-MeanForce, Campisi-MeanForce-ParitionFunction, Miller-Anders-MeanForce, Rivas-nonEquilibMeanForce}. In addition, there are \textit{semiclassical} approaches where a coarse-grained version of the system state in the energy eigenbasis and semiclassical definitions of heat and work have been employed \cite{Bender_2000, Abe, Polkovnikov-diag-entr}.

Despite extensive efforts, some of the existing approaches neither are consistent with each other nor do they always reproduce expected results according to thermodynamics. Although there exist attempts to overcome such inconsistencies through ad hoc methods \cite{Shi-heatCoherence,Su-heatCoherence,Rivas-nonEquilibMeanForce}, the issue remains unresolved.  

To further clarify the issue, we  note the following drawbacks of the conventional framework. (i) Its formulation has been originally devised only for particular conditions, e.g., a Markovian dynamics with a constant or relatively slow-varying Hamiltonian, when the dissipative part does not explicitly depend on the physical system Hamiltonian. Despite this subtlety, the conventional definitions of heat and work have been applied almost arbitrarily in the literature to more general scenarios. In fact, dissipative part of a master equation may explicitly depend on the physical Hamiltonian of the system \cite{Dann-Kosloff-STequilibrium, Nafari}. Hence in such physical scenarios, external driving may also contribute to the heat exchanged. This aspect introduces further ambiguities and hinders a clear and thermodynamically consistent assignment of the heat and work concepts. (ii) Since the description of the dynamics in the form of the conventional Lindblad master equation is not unique \cite{Funo19SpeedOpen}, different energy values can be assigned to differently chosen generators of the same dynamics. This indicates that part of the energy assigned to the dissipative part of the master equation may not necessarily lead to an entropy change, and should be identified as work. Specifically, for initial states of the decoherence-free subspace type \cite{Lidar-Whaley-DFS,Blume-Kohout-DFS}, the dynamics leads to no dissipation along the corresponding trajectories and all the energy change, if any, would be of the work type. 
%(iii) Although in the \textcolor{blue}{conventional} framework it is only the dissipative part of the dynamics that contributes to entropy change, as we argue later, not the whole dissipative part indeed yields entropy change---and hence heat change. This necessitates a rigorous analysis to partition the dissipative part to terms yielding entropy (heat) change and terms lacking such a property. 
Recently, a scheme to partially remedy the shortcomings of the conventional definitions
has been proposed in Ref. \cite{Kurizki-ergo}, where heat has been defined by subtracting \textit{ergotropy} (maximum extractable work from a system in a cyclic unitary process) \cite{Allahverdyan-ergotropy} from the dissipative part of the internal energy change by considering a \textit{virtual} instantaneous unitary transformation. Nevertheless, an extension of ergotropy to open systems is still elusive, and since this extra transformation does not rely on the real physical dynamics, trajectory-dependent quantities such as heat can be fictitious. For an alternative approach, separating the internal energy based on conserved quantities of the dynamics, see Ref. \cite{Manzano}.

In this Letter, on pure thermodynamic grounds supplemented with dynamical and geometric arguments, we circumvent the issue and ambiguities discussed above and put forward a set of universal definitions for heat, work, and entropy production based explicitly on trajectories in the state space of a quantum system (note the difference with quantum jump \textit{trajectories} arising in stochastic dynamics of monitored systems \cite{Horowitz, Lutz-trajectory}). Our framework is general and independent of the dynamics which generates the trajectory, and it is thus applicable to any time-continuous evolution.

In particular, following standard thermodynamics wherein energy change in an isentropic process is identified as work \cite{book:Callen, Weimer-Mahler-WorkHeat}, we define heat solely as the part of the internal energy of the system which can be associated with entropy change. While the first law of thermodynamics treats heat and work on an equal footing, the second law distinguishes them. Essentially heat is a form of disorganized energy, and some disorganization (entropy) will flow with it \cite{book:Cengel-thermodynamics-engineering}. Hence, heat is introduced as the part of the energy change which can be accompanied by an entropy change, whereas the energy transfer in the form of work definitely does not lead to any entropy change. We use this key feature to identify heat and work and accordingly obtain a computable expression for the irreversible entropy production. We show that this quantity varies due to the change in variables of the system other than its energy. This clarifies why in general scenarios heat and entropy changes are not necessarily monotonic with respect to each other.

Consider the evolution of the system in a time interval $t\in[0,t_{f}]$ described by a time-dependent density matrix $\varrho(t)$, expressed in its spectral decomposition as
\begin{align}
\varrho(t)=\textstyle{\sum_{k=1}^{D}} r_{k} (t)| r_{k} (t)\rangle\langle r_{k} (t)|.
\label{rho-spect}
\end{align}
One can consider $\{\varrho(t)\}$ as a \textit{trajectory} in the state space of the system, starting from given $\{\varrho(0)\}$ (for brevity, we drop all time dependence henceforth unless necessary). 
Note that the change in $\varrho$ can be decomposed as $d\varrho=\dbar \varrho^{(\mathrm{ev})}+ \dbar \varrho^{\mathrm{(ep)}}$, where $\dbar \varrho^{(\mathrm{ev})}=\textstyle{\sum_{k}} dr_{k} | r_{k}\rangle\langle r_{k}|$ is the change due to the variation of the eigenvalues and $\dbar \varrho^{(\mathrm{ep})}=\textstyle{\sum_{k}} r_{k} (|dr_{k}\rangle\langle r_{k}|+|r_{k}\rangle\langle dr_{k}|)$ is due to the variation in the eigenprojectors, where $\dbar$ denotes inexact differential. In addition, we observe that the entropy $S(\varrho)=-\mathrm{Tr}[\varrho\ln\varrho]$ changes only when the eigenvalues vary,
\begin{align}
dS= -{\textstyle{\sum_{k}}} dr_{k} \ln r_{k} = -\mathrm{Tr}[\dbar \varrho^{(\mathrm{ev})} \ln \varrho].
\label{ent}
\end{align}
In a system with the physical Hamiltonian $H$, the system internal energy $U=\mathrm{Tr}[\varrho H]$ changes along the trajectory with $t\to t+dt$ as $U\to U=U+dU$, with $d{U} = \mathrm{Tr}[d{\varrho} \,H]+\mathrm{Tr}[\varrho\, d{H}]$. Here,  the contribution associated with $\dbar \varrho^{(\mathrm{ev})}$ is solely related to the change of the eigenvalues, which is in line with how the entropy changes. Hence, we define the heat change as
\begin{equation}
\dbar\mathbbmss{Q} =\mathrm{Tr}[\dbar \varrho^{(\mathrm{ev})} H] = \textstyle{\sum_{k}} dr_{k} \langle  r_{k} |H| r_{k}  \rangle, 
\label{def:heat}
\end{equation}
and assign the remaining variations to work change, 
\begin{align}
\dbar\mathbbmss{W} = &\mathrm{Tr}[\varrho\, dH] + \mathrm{Tr}[\dbar\varrho^{(\mathrm{ep})} H] \nonumber\\
= & \textstyle{\sum_{k}}  r_{k}   \big(\langle r_{k}|dH| r_{k} \rangle + \langle dr_{k}|H| r_{k} \rangle + \langle  r_{k} |H| dr_{k} \rangle\big), 
\label{def:work}
\end{align}
such that
\begin{equation}
dU = \dbar\mathbbmss{Q} + \dbar\mathbbmss{W}
\label{1st-law}
\end{equation}
encompasses the first law of thermodynamics.

%%%%%%%%%%%%%%%%%%%%%%%%%%%%%%%%%%%%%%%%%%%%%%%%%%%%%%%%%%%%%%%%
\textit{Dynamical analysis of the heat and work definitions}.---We use a recently proposed \textit{trajectory-based shortcut to adiabaticity} (TB-STA) framework for open-system dynamics, which identifies a particular dynamical equation of motion that generates any given time-dependent density matrix or trajectory \cite{Alipour:STA}. This technique, based on the general Lindblad-like equation described later, is not limited by the system-environment coupling strength or initial correlation. Since the TB-STA equation describes a given trajectory, its coherent and dissipative parts are directly related to the coherences and dissipation in the course of the evolution, which contrast with the conventional Markovian Lindblad master equation. Our main result shows that, among all possible dynamical equations describing a given trajectory, TB-STA allows us to unambiguously separate the change of the internal energy into heat and work for any open system. We prove that energy change assigned to the (so-called) dissipative part of the equation is accompanied by entropy change, and it is therefore \textit{heat}. The coherent part, which does not involve entropy change, corresponds to a \textit{dissipative work}, 
%which properly accounts for the work done on the system through its interaction with the environment 
which for its relation to nonunitarity of the dynamics entails implicit existence of some ambient environment.
 
Having a trajectory $\{\varrho(t)\}$ at hand, an associated differential equation describing its dynamics is given by the following Lindblad-like equation \cite{Alipour:STA}:
\begin{align}
%\dot{\varrho}=-i[\mathbbmss{H}_{\textsc{cd}},\varrho] +\mathbbmss{D}_{\textsc{cd}}[\varrho],
\dot{\varrho}=-i[\mathbbmss{h},\varrho] +\mathbbmss{D}_{\textsc{cd}}[\varrho],
\label{geom}
\end{align}
where 
\begin{gather}
%\mathbbmss{H}_{\textsc{cd}}=\mathbbmss{H}+\mathbbmss{h}, \label{H-cd}\\
\mathbbmss{h} =  i \textstyle{\sum_{k}}\left( | \dot{r}_{k} \rangle\langle r_{k} |-\langle  r_{k} | \dot{r}_{k} \rangle | r_{k} \rangle\langle    r_{k} |\right),
\label{H_{1}}\\
\mathbbmss{D}_{\textsc{cd}}[\varrho]=\textstyle{\sum_{kj}} \mathbbmss{c}_{kj}\big(\mathbbmss{L}_{kj}\varrho \mathbbmss{L}_{kj}^\dag - \frac{1}{2}\{\mathbbmss{L}_{kj}^\dag \mathbbmss{L}_{kj},\varrho\}\big), \label{Diss}\\
\mathbbmss{L}_{kj}=|r_{k} \rangle \langle r_{j}|,~~ \mathbbmss{c}_{kj}= (1-\delta_{r_{j}0})\dot{r}_{k}/(Dr_{j}),
\label{rates}
\end{gather}
with the dot denoting time differentiation. Several remarks are in order: (i) %$\mathbbmss{H}_{\textsc{cd}}$ is a \textit{counterdiabatic} (CD) Hamiltonian associated with $\mathbbmss{H}$ \cite{Alipour:STA}, with the extra 
$\mathbbmss{h}$ operates as a \textit{counterdiabatic} (CD) Hamiltonian which alone generates a parallel transport $|r_{k}(0)\rangle\to |r_{k}(t)\rangle\,\forall k$ (cf. CD Hamiltonians in the energy sense \cite{Demirplak03,Berry09}); (ii) in the CD dissipator $\mathbbmss{D}_{\textsc{cd}}$ the anticommutator identically vanishes, whence the dissipator reduces exclusively to jumps in the instantaneous eigenbasis of $\varrho$; (iii) the dynamical equation (\ref{geom}) is universal, irrespective of any physical setting (system Hamiltonian and the environment) in which the system has obtained this trajectory. In the standard technical sense \cite{BreuerBook,book:Rivas-Huelga}, this dynamical equation for a trajectory is different from the master equation of the system. The same physical system prepared in a different initial condition with the same physical setting can yield a different trajectory---hence a different dynamical equation associated with the new trajectory; and (iv) a Lindblad-like master equation for general open-system dynamics has also been introduced recently \cite{ULL}.

%%%%%%%%%%%%%%%%%%%%%%%%%%%%%%%%%%%%%%%%%%%%%%%%%%%%%%%%%%%%%%%%
\textit{Comparison with the conventional definitions}.---Along the trajectory, the conventional definitions are $Q=\int_{0}^{t}dt\,\mathrm{Tr}[\dot{\varrho} H]$ for heat change and $W=\int_{0}^{t}dt\,\mathrm{Tr}[\varrho \dot{H}]$ for work exchange, hence $d{U}=\dbar{Q} + \dbar{W}$ [cf. Eq. \eqref{1st-law}].  
\ignore{
%%%%%%%%%%%%%%%%%%%%%%%%%%%%%%%%%%%%%%%%%%%%%%%%%%%%%%%%%%%%%%%%
\begin{figure}[t]
\includegraphics[width=0.9\linewidth]{fig-1}
\caption{Schematic representation of different parts of internal energy change in dissipative evolution of the system density matrix $\varrho$. Given a trajectory $\{\varrho(t)\}$ for a system that is in contact with a bath, part of the energy change in the system is in the form of heat $d\mathbbmss{Q}$. In addition to the work done by or on the system due to driving its physical Hamiltonian $dW$, there is another part, dissipative work $d\mathbbmss{W}_{\textsc{cd}}$, which is performed to keep the system on the given trajectory.}
\label{Wcd}
\end{figure}
%%%%%%%%%%%%%%%%%%%%%%%%%%%%%%%%%%%%%%%%%%%%%%%%%%%%%%%%%%%%%%%%
}
Now rewrite $\dbar Q$ as
\begin{equation}
\dbar Q = \textstyle{\sum_{k}}  dr_{k} \langle  r_{k} |H| r_{k}  \rangle +  \textstyle{\sum_{k}}  r_{k}   \big(\langle  dr_{k}|H| r_{k} \rangle+\langle  r_{k} |H| dr_{k} \rangle\big).
\label{dQ-dWcd}
\end{equation}
The first sum originates from a change in the eigenvalues of the state, contributing when not all $dr_{k}$s vanish (similar to the reason for the change in $S$);   
%\textcolor{blue}{Note that from Eq. \eqref{ent} the latter is exactly the sufficient condition for having a change in the entropy}.
%In other words, the time variations of the first term in $dQ$ and $dS$ are concomitant. 
the second sum is solely associated with the change of the eigenvectors. 
%, whereas the first term involves the change of the eigenvalues (similar to the reason for the change in $S$). 
The first sum can be rewritten as $\mathrm{Tr}\big[\mathbbmss{D}_{\textsc{cd}}[\varrho]H\big] dt$, which is related to the \textit{dissipative} part of the trajectory dynamical equation (\ref{geom}); whereas the second sum can be recast as $-i\mathrm{Tr}\big[[H,\mathbbmss{h}] \varrho\big]dt$, which is associated with the \textit{coherent} part of the dynamical equation---hence dissipative \textit{work}. From this perspective, we can consider the following relations as a justification of our definitions:
%These observations about the separation in $dQ$ prompt us to introduce the following modified definitions for the (infinitesimal) exchanged heat and work:
\begin{align}
{\dbar}\mathbbmss{Q}  &= {\dbar} Q - {\dbar}\mathbbmss{W}_{\textsc{cd}} =\mathrm{Tr}\big[\mathbbmss{D}_{\textsc{cd}}[\varrho]H\big] dt, \label{dQ}
\\
{\dbar}\mathbbmss{W} &= {\dbar} W+ {\dbar} \mathbbmss{W}_{\textsc{cd}} =  \mathrm{Tr}\big[\varrho\big(\dot{H} -i [H,\mathbbmss{h}]\big)\big] dt, \label{dWtot}
\end{align}
where $\dbar\mathbbmss{W}_{\textsc{cd}}$ is an environment-induced dissipative work, 
\begin{align}
\dbar\mathbbmss{W}_{\textsc{cd}}&=\textstyle{\sum_{k}}  r_{k} \big(\langle dr_{k}|H| r_{k} \rangle+\langle  r_{k} |H| dr_{k} \rangle\big) \label{dWcd}\\
& = -i \mathrm{Tr}\big[[H,\mathbbmss{h}]\varrho\big]d t, \nonumber
\end{align}
and is generated due to the CD evolution. In this framework the work is due to both driving the system through varying its physical Hamiltonian ($\dbar{W}$) and the CD evolution along the trajectory due to the environment ($\dbar \mathbbmss{W}_{\textsc{cd}}$). Importantly, unlike in Ref. \cite{Demirplak03}, the CD term here is not related to an external physical CD control of the Hamiltonian; rather, it corresponds to the natural CD evolution along the state trajectory. %---Fig. \ref{Wcd}. 
%In this sense, we can say that the CD work prevents the system from deviating from its trajectory. 
%We also recover the first law of thermodynamics [Eq. (\ref{1st-law})].

%%%%%%%%%%%%%%%%%%%%%%%%%%%%%%%%%%%%%%%%%%%%%%%%%%%%%%%%%%%%%%%%
\textit{Relation to the semiclassical definitions.---}In the \textit{semiclassical} formulation of thermodynamics for quantum systems, heat and work are defined differently \cite{Polkovnikov-diag-entr}. Using the instantaneous eigenbasis of the system Hamiltonian $H=\sum_{n}E_{n} |E_{n}\rangle \langle E_{n}|$,
%Assuming the instantaneous spectral decomposition of the physical Hamiltonian of the system  and using this energy basis, 
%the internal energy $U$ becomes the \textit{average} energy 
we obtain $U=\sum_{n} p_{n} E_{n}$, where $p_{n}=\langle E_{n}|\varrho|E_{n} \rangle$ is the population of the energy eigenstate $|E_{n}\rangle$. From the identity
%\begin{align}
$dU = \textstyle{\sum_{n}} dp_{n}\, E_{n} +\textstyle{\sum_{n}} p_{n}\, dE_{n}$, 
%\end{align}
one can read the semiclassical heat and work variations as 
\begin{gather}
\dbar{q}= \textstyle{\sum_{n}} dp_{n}\, E_{n},\\
\dbar{w}= \textstyle{\sum_{n}} p_{n}\, dE_{n}.
\end{gather}
However, if instead of using  the instantaneous eigenbasis of $H$, we evaluate the trace in $U$ in the instantaneous eigenbasis of $\varrho$ [Eq. (\ref{rho-spect})], then ${U}=\sum_{k}  r_{k} H_{k}$ and
%\begin{align}
$d{U} = \textstyle{\sum_{k}} dr_{k}\,H_{k} + \textstyle{\sum_{k}} r_{k}\,dH_{k}$,  
%\end{align}
where $H_{k}=\langle  r_{k} | H| r_{k}  \rangle$. As in the semiclassical setting, one can now read the first term as the heat change and the second one as the work change. We observe that these are equivalent to the TB-STA definitions [Eqs. \eqref{dQ} and \eqref{dWtot}], 
%\begin{gather}
$\dbar\mathbbmss{Q}= \textstyle{\sum_{k}} dr_{k}\,  H_{k}$ and 
$\dbar\mathbbmss{W}= \textstyle{\sum_{k}}  r_{k} \, dH_{k}$. 
%\end{gather}
One can argue that the instantaneous eigenbasis of $\varrho$ is preferred in calculating the energy contributions \cite{Basis}. A closer inspection of $\dbar{q}$ reveals that it includes the energy (not heat) change $\dbar\mathbbmss{W}_{\textsc{cd}}$ which we have assigned to the CD Hamiltonian in the form of work, 
\begin{align}
\hskip-0.2mm\dbar{q}=\dbar\mathbbmss{Q} +\dbar \mathbbmss{W}_{\textsc{cd}}{+}\textstyle{\sum_{n}}  E_{n} \big(\langle dE_{n}|\varrho |E_{n} \rangle + \langle E_{n} |\varrho |dE_{n}\rangle \big).
\end{align}
Thus, since $\dbar q$ has contributions from both heat and work exchanges (in our sense), we conclude that in general the semiclassical definitions of heat and work fail to properly account for various contributions to the internal energy change. 
%\textcolor{red}{AUR: do you conclude this because the 3 terms in the equation above are mixed between heat, work, and something else, and not just heat? can this statement be more specific? maybe add "at least compared to our proposition" at the end?}

%%%%%%%%%%%%%%%%%%%%%%%%%%%%%%%%%%%%%%%%%%%%%%%%%%%%%%%%%%%%%%%%
\textit{Irreversible/Internal entropy production.}---The entropy change $dS$ [Eq. (\ref{ent})], together with the heat change $\dbar \mathbbmss{Q}$ [Eq. \eqref{dQ}] give the change in the irreversible entropy $\mathbbmss{S}$ as 
\begin{align}
\dbar \mathbbmss{S}\equiv dS - \beta\, \dbar\mathbbmss{Q}=\textstyle{\sum_{k}} dr_{k} \langle  r_{k} | (\mathbbmss{H}-\beta H)| r_{k} \rangle,
\label{dSp}
\end{align} 
where $\mathbbmss{H} = - \log \varrho$ %(or $\varrho=e^{-\mathbbmss{H}}$)} 
and $\beta$ is
%which depends on the difference between the virtual and the real Hamiltonians of the system, and 
the nonequilibrium, instantaneous inverse temperature of the system given by (if $k_{B}\equiv 1$) \cite{qtemperature} 
%based on the partial derivative of the entropy with respect the system energy, given by
\begin{align}
\beta=\left(\partial S/\partial U\right)_{x_{2},x_{3},\cdots} = \mathrm{Cov}{(H, \mathbbmss{H}})/\mathrm{Cov}(H,H).
\label{qT}
\end{align}
Here $\{x_{i}\}$ are a set of independent variables obtained from the expectation values of a complete set of traceless orthonormal observables $\{O_{i}\}_{i=2}^{D^{2}-1}$, with $O_{0}=\openone/\sqrt{D}$ (the normalized identity operator of dimension $D$) and $O_{1}=(H-\mathrm{Tr}[H]/D)/\sqrt{D\,\mathrm{Cov}(H,H)}$. In addition, 
%$\mathrm{Cov}_{\mathrm{mc}}(X, Y)=\mathrm{Tr}[XY\varrho_{\mathrm{mc}}]-\mathrm{Tr}[X\varrho_{\mathrm{mc}}]\,\mathrm{Tr}[Y\varrho_{\mathrm{mc}}]$ 
$\mathrm{Cov}(X, Y)\equiv \mathrm{Tr}[X Y]/D -\mathrm{Tr}[X]\,\mathrm{Tr}[Y]/D^{2}$. 
%, where $\varrho_{\mathrm{mc}}=\openone/D$ is the microcanonical state. 
In terms of the $\{x_{i}\}$ variables we have $\dbar\varrho^{(\mathrm{ev})} = [D\, \mathrm{Cov}(H,H)]^{-1/2} \dbar\mathbbmss{Q}\, H + \sum_{i\geqslant2} {\dbar x}_{i}\, O_{i}$, where $\dbar x_{i}=\mathrm{Tr}[\dbar \varrho^{(\mathrm{ev})}\, O_{i}]$. 
%---we have used the projection over the operator basis $\varrho = \sum_{i=0}^{D^{2}-1}x_{i} O_{i}$ here. 
As a result, we obtain $dS=\beta\dbar\mathbbmss{Q}+\sum_{i\geqslant 2} \mathrm{Tr}[O_{i} \mathbbmss{H}]\,\dbar x_{i}$, whence %an equivalent expression for $\dbar \mathbbmss{S}$ is obtained as 
\begin{align}
\dbar \mathbbmss{S} 
%&\equiv  dS-\beta d\mathbbmss{Q}\nonumber\\
=\textstyle{\sum_{i\geqslant 2}}\mathrm{Tr}[O_{i} \mathbbmss{H}]\,\dbar x_{i}, 
%\nonumber\\
%&\equiv\textstyle{\sum_{i\geqslant 2}}\mathrm{Tr}[O_{i} \mathbbmss{H}]\,\mathrm{Tr}[O_{i}\delta \varrho^{(\mathrm{ev})}].
\label{dSp-observable}
\end{align} 
which is indeed independent of the choice of $\{x_{i}\}$ \cite{SM}. We observe that $\dbar \mathbbmss{S}$ is determined only in terms of the system variables, independent of the existence of an environment in a thermal equilibrium, and the system Hamiltonian does not play any explicit role therein; it is an irrelevant observable---see Ref. \cite{SM} for further discussion. 
%irreversible entropy can be produced in the system due to the change in any variable (among the set of introduced variables) of the system other than the energy. 
%in which $\{O_{i}\}$ are a set of orthonormal, traceless, and Hermitian operators that together with the identity and the normalized traceless version of the system Hamiltonian form a set of complete bases (for more details see SM).  
We remark that, besides using $\dbar\mathbbmss{Q}$ in our definition of irreversible entropy production (\ref{dSp}), unlike the earlier literature \cite{Paternostro-entropy-production} here $\beta$ is associated to the system (not to a large environment or heat bath). A discussion of why $\beta$ is associated with the system can be found in Supplemental Material \cite{SM}. Another alternative expression for $\dbar \mathbbmss{S}$ can be obtained in terms of the relative entropy $S(\varrho\|\varrho_{\mathrm{eq}}) = \mathrm{Tr}[\varrho\ln \varrho-\varrho\ln \varrho_{\mathrm{eq}}]$ between the state $\varrho$ and the instantaneous canonical Gibbs state $\varrho_{\mathrm{eq}}\equiv e^{-\beta H}/\mathrm{Tr}[e^{-\beta H}]$ as \cite{SM} 
%\begin{align}
$\dbar \mathbbmss{S} =-dS(\varrho\|\varrho_{\mathrm{eq}}) +\beta \dbar\mathbbmss{W}_{\textsc{cd}} + \mathrm{Tr}[(\varrho-\varrho_{\mathrm{eq}})d(\beta H)]$.
%\nonumber\\&+d\beta\,\mathrm{Tr}[(\varrho-\varrho_{\mathrm{eq}}) H].
%\label{S-ir}
%\end{align} 
%By using $dS(\varrho\|\varrho_{\mathrm{eq}})\leqslant 0$ this relation can be turned into an inequality, which should be contrasted with Refs. \cite{esposito-EntropyProduction, Uzdin}. 
%It is straightforward to show that under the conditions of Ref. \cite{esposito-EntropyProduction}, Eq. (\ref{S-ir}) reproduces the same results.
We also note that interestingly from Eq. (\ref{dSp-observable}) an inequality can also be derived as \cite{SM}
\begin{equation}
\dbar \mathbbmss{S} + b(\mathbbmss{O})\,\dbar B\geqslant 0,  
\label{Clausius-ineq}
\end{equation}
which is reminiscent of a generalized Clausius inequality \cite{Uzdin, esposito-EntropyProduction} with all quantities depending explicitly on the system. Here $b(\mathbbmss{O})$ is the spread of $\mathbbmss{O}=\sum_{i\geqslant 2}O_{i}\otimes O_{i}$ and $\dbar B=-\sum_{k}\ln r_{k}\sum_{k'|dr_{k'}\geqslant 0}dr_{k'}$. 

For dynamically \textit{closed quantum systems} (where $\dot{\varrho}=-i[H,\varrho]$, assuming $\hbar \equiv 1$), we have $\dot{r}_{k}=0\, \forall k$, hence $dU=\dbar \mathbbmss{W} = \dbar W$ and $\dbar \mathbbmss{S}= dS =\dbar Q = \dbar \mathbbmss{Q}= \dbar \mathbbmss{W}_{\textsc{cd}} = 0$. This case also naturally includes phase-space-preserving cooling processes \cite{Chen10}. For dynamically \textit{open quantum systems} weakly coupled to a large environment, when $H$ is constant or slowly driven the dynamics obeys a Markovian Lindblad master equation $\dot{\varrho}= \mathpzc{L}[\varrho]$, where $\mathpzc{L}[\varrho]=-i[H+H_{\mathrm{Lamb}},\varrho]+ \mathpzc{D}[\varrho]$, with $H_{\mathrm{Lamb}}$ being the environment-induced Lamb-shift correction and $\mathpzc{D}[\varrho] = \sum_{\alpha} c_{\alpha}\big(L_{\alpha}\varrho L^{\dag}_{\alpha}-(1/2)\{L^{\dag}_{\alpha}L_{\alpha},\varrho \} \big)$ ($c_{\alpha}>0$) being the quantum dissipator \cite{BreuerBook, book:Rivas-Huelga} [cf Eq. \eqref{Diss}]. Additionally, if the system starts at $\varrho_{\mathrm{eq}}(0)$ and also $\varrho_{\mathrm{eq}}(t)$ is the unique instantaneous steady state of the dynamics \cite{Spohn-relaxing, Bach}, one can prove that $\varrho(t)\approx \varrho_{\mathrm{eq}}(t)\,\forall t$ \cite{Cavina}, namely the trajectory remains near the steady state, and also $dS-\beta \, \dbar Q\geqslant 0 $ \cite{Spohn-EP, Alicki, SciRep:Alipour-corr}. This implies that $\dbar \mathbbmss{S} - \beta\, \dbar\mathbbmss{W}_{\textsc{cd}}\geqslant 0$, where $\dbar\mathbbmss{W}_{\textsc{cd}}$ [Eq. (\ref{dWcd})] is an energetic cost associated to how different the real trajectory is from the quasistatic one---cf. the general relation (\ref{Clausius-ineq}). This relation can be considered as a manifestation of the Clausius inequality \cite{esposito-EntropyProduction, Uzdin}. Note that in general $\dbar\mathbbmss{W} \neq \dbar W$ and $\dbar\mathbbmss{Q}\neq \dbar Q =\mathrm{Tr}[\mathpzc{D}[\varrho] H]dt$ [cf. Eq. (\ref{dQ})]; whereas in the \textit{quasistatic} regime $\varrho(t) = \varrho_{\mathrm{eq}}(t)$, we have $|r_{k}\rangle = |E_{k}\rangle$, $\mathbbmss{H} = \beta H$, and thus $\dbar \mathbbmss{W}_{\textsc{cd}} = 0$, $\dbar \mathbbmss{Q} = \dbar Q$, $\dbar \mathbbmss{S} = 0$, and $dS = \beta \dbar Q$. The latter is a manifestation of the thermodynamic \textit{reversibility} \cite{book:Callen}. 
 
%%%%%%%%%%%%%%%%%%%%%%%%%%%%%%%%%%%%%%%%%%%%%%%%%%%%%%%%%%%%%%%%
\ignore{
(ii) It has also been shown in Refs. \cite{esposito-EntropyProduction,Paternostro-entropy-production} that under some conditions such as weak-coupling, and a large environment with constant Hamiltonian, when the whole energy change of the environment can be considered as heat and the whole heat is transferred to the system (this implies also a weak system-environment correlation assumption \cite{SciRep:Alipour-corr}), by assuming that the initial state of the bath is an equilibrium state $\varrho_{\mathsf{B}}^{\mathrm{eq}}$ with initial inverse temperature $\beta$, then $\Delta S + \beta \Delta Q_{B}=S(\varrho_{\mathsf{SB}}\|\varrho_{\mathsf{S}}\otimes \varrho^{\mathrm{eq}}_{\mathsf{B}})$. Since relative entropy is a positive quantity, this relation has been considered as a manifestation of the second law of thermodynamics. It is evident that under the same assumptions $d\mathbbmss{S}$ also reduces to the same relation and is positive. This can be seen assuming that the whole energy transfer with the environment is in the form of heat, which leads to $d\mathbbmss{W}_{\textsc{cd}}=0$ and thus $d\mathbbmss{Q}=dQ$. In addition, the large environment assumption yields that the inverse temperature of the system is almost equal to that of the environment \cite{qtemperature}. Hence $d\mathbbmss{S}$ becomes equival to $dS-\beta dQ$ and is then positive.
}
\ignore{
%%%%%%%%%%%%%%%%%%%%%%%%%%%%%%%%%%%%%%%%%%%%%%%%%%%%%%%%%%%%%%%%
\textit{Relation to balanced loss and gain.---}One can describe dynamics of the system, given in Eq. \eqref{geom}, equivalently through a balanced gain and loss formalism  \cite{Alipour:STA}. Within this context, where the dynamics is described by a non-Hermitian generator, evolution of any density matrix can be decomposed into two parts: a coherent part which is similar to the coherent part of Eq. \eqref{geom} and related to the unitary rotation of the eigenvectors of the density matrix, and a dissipative part which is related to the changes of the eigenvalues and can account for gain and loss. This decomposition can be readily seen by looking at the generator of the dynamics which is given by a non-Hermitian operator
\begin{equation}
H_{\textsc{gl}}=\mathbbmss{H}_{\textsc{cd}}+i \Gamma,
\end{equation}
such that 
\begin{align}
\Gamma= - \frac{1}{2}\textstyle{\sum_{k}}\big(\dot{r}_{k}/r_{k}\big) |r_{k} \rangle\langle r_{k}|,
\end{align}
is Hermitian, and the dynamical equation can be recast as 
\begin{equation}
\dot{\varrho}=-i[\hskip-0.9mm[ H_{\textsc{gl}},\varrho ]\hskip-0.9mm],
\end{equation}
where $[\hskip-0.9mm[ A,B ]\hskip-0.9mm]=A B -B^{\dag} A^{\dag}$. Here $i \Gamma$ is the generator of heat production or absorption processes, and $\mathbbmss{H}_{\textsc{cd}}$ is the generator of work production or absorption processes. Using this formalism, the heat production rate is given by 
\begin{align}
\dot{\mathbbmss{Q}}=-\mathrm{Tr}[\{H,\Gamma\}\varrho],
\label{dQ-GL}
\end{align}
which is equivalent to Eq. \eqref{dQ}. Using Eqs. \eqref{dWtot} and \eqref{dQ-GL} one can assign a Hermitian operator $R_{\mathbbmss{W}}=\dot{H}-i[H,\mathbbmss{h}]$ to the work rate such that $\dot{\mathbbmss{W}}=\mathrm{Tr}[R_{\mathbbmss{W}}\varrho]$, and a Hermitian operator $R_{\mathbbmss{Q}}=-\{H,\Gamma\}$ to the heat rate such that $\dot{\mathbbmss{Q}}=\mathrm{Tr}[R_{\mathbbmss{Q}}\varrho]$. 
}
%%%%%%%%%%%%%%%%%%%%%%%%%%%%%%%%%%%%%%%%%%%%%%%%%%%%%%%%%%%%%%%%
\begin{figure}[tp]
\includegraphics[width=0.7\linewidth]{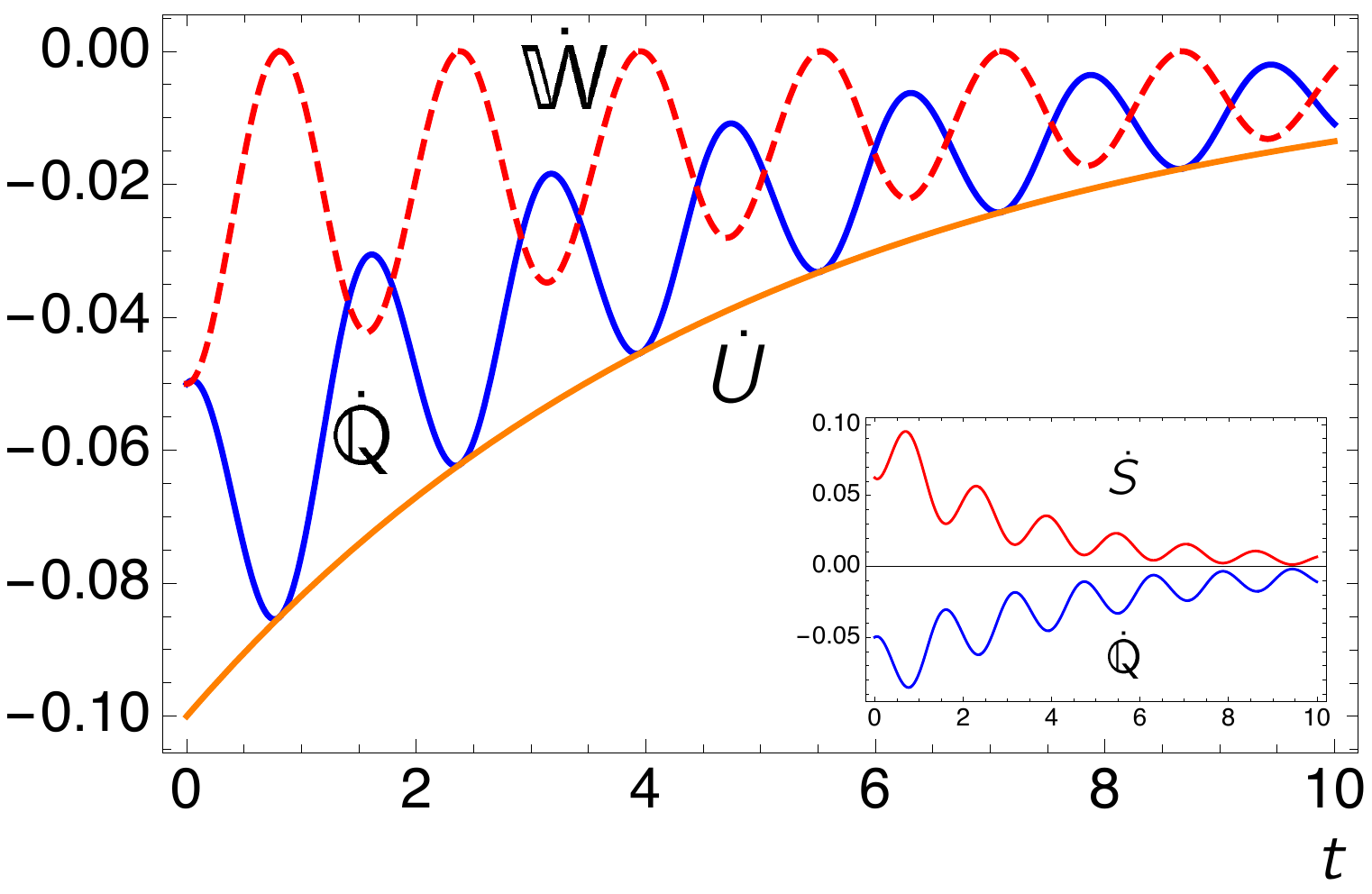}
\caption{Rates of internal energy (solid orange), heat (solid blue), and work (dashed red) changes vs. time (in natural units $\hbar\equiv k_{B}\equiv 1$), for a qubit in a heat bath as given in example I, with $\gamma=0.1$ and $\omega_{0}=1$. The integrated values are $U =-0.25$, $\mathbbmss{Q}=-0.138$, and $\mathbbmss{W}=-0.112$, respectively, where $X=\int_{0}^{10}dt\,\dot{X}$ with $X\in\{U,\mathbbmss{Q},\mathbbmss{W}\}$. Inset compares $\dot{S}$ and $\dot{\mathbbmss{Q}}$.}
\label{figure2}
\end{figure}
%%%%%%%%%%%%%%%%%%%%%%%%%%%%%%%%%%%%%%%%%%%%%%%%%%%%%%%%%%%%%%%%

%%%%%%%%%%%%%%%%%%%%%%%%%%%%%%%%%%%%%%%%%%%%%%%%%%%%%%%%%%%%%%%%
\textit{Example I: Qubit in a Markovian environment.---}Consider a qubit with $H=\omega_{0} \sigma_{z}$ that weakly interacts with a Markovian environment such that $\dot{\varrho}=-i[\omega_{0} \sigma_{z},\varrho]+\gamma \big(\sigma_{x} \varrho \sigma_{x}-\varrho\big)$, where $\omega_{0}$ and $\gamma$ are positive constants, and $\sigma_{x}=|0\rangle\langle 1| + |1\rangle\langle 0|$ and $\sigma_{z}=|0\rangle \langle 0| - |1\rangle \langle 1|$ are the $x$- and $z$-Pauli operators \cite{BreuerBook}. The quantity $\dot{U}=\mathrm{Tr}[H \dot{\varrho}]=-2\gamma \omega_{0} \mathrm{Tr}[\sigma_{z} \varrho]$ is fully due to the dissipative part of the dynamics. We consider three different initial states (for details, see Ref. \cite{SM}). (i) Thermal state $\varrho_{\mathrm{eq}}(0)$,  
%e^{-\beta\omega_{0}\sigma_{z}}/\mathrm{Tr}[e^{-\beta\omega_{0}\sigma_{z}}]$, 
with $\beta(0)$ being the initial inverse temperature of the system (this can differ from that of the environment, and the latter need not be in a thermal state). In this case, we obtain $\dot{\mathbbmss{Q}}=\dot{U}= 2\gamma \omega_{0} e^{-2\gamma t} \tanh [\beta(0) \omega_{0}]$, $\dot{q}=\dot{\mathbbmss{Q}}$, and $\dot{w}=\dot{\mathbbmss{W}}$. (ii)  Pure state $|\psi(0)\rangle=(|0\rangle+|1\rangle)/\sqrt{2}$. One can see that $\dot{U}=\dot{\mathbbmss{Q}}= \dot{\mathbbmss{W}}=\dot{q}=\dot{w}=0$. (iii) Starting from $\varrho(0)=(\openone+[\sigma_{x}+\sigma_{z}]/2)/2$, we can obtain the heat, work, and internal energy change in the system---Fig. \ref{figure2}. It can be shown that $\dot{q}=\dot{U}=-(\gamma/2) e^{-2 \gamma t}$ and $\dot{w}=0$. 

%%%%%%%%%%%%%%%%%%%%%%%%%%%%%%%%%%%%%%%%%%%%%%%%%%%%%%%%%%%%%%%%
\textit{Example II: Qubit in a dephasing environment.---}Consider the dephasing master equation for a qubit, $\dot{\varrho}=-i[\omega_{0} \sigma_{z},\varrho]+\gamma (\sigma_{z} \varrho \sigma_{z}-\varrho)$. Generally, in this process the system energy is preserved and there is no net energy exchange between the system and the environment. However, the system and the environment can still exchange heat and work. This can be contrasted with Landauer's principle, where the environment does work to erase phase information and the system releases heat back into the environment. Let $\omega_{0}=1$ and $\gamma=0.1$ and consider two different cases: (i) The initial state is $|\psi(0)\rangle=(|0\rangle +\sqrt{2}|1\rangle)/\sqrt{3}$: we obtain that $\dot{\mathbbmss{Q}}=-\dot{\mathbbmss{W}}_{\textsc{cd}}=(8/15)/[8 +e^{2 t/5}]$ and $\dot{S}=4 \ln [(1+2\Delta)/(1-2\Delta)]/(15 e^{t/5} \sqrt{8 + e^{2t/5}})$, where $\Delta=(1/6) e^{-t/5}\sqrt{8 + e^{2 t/5}}$. This process can be compared with an isothermal process in ideal gases, where the energy is constant and the heat exchanged is equal to the work exchanged with the opposite sign. However, since this dynamics does not exactly model an ideal gas (its internal energy is not proportional to its temperature), $\beta$ varies and thus the process is not isothermal. (ii) The initial state is $|\psi(0)\rangle=(|0\rangle +|1\rangle)/\sqrt{2}$: in this case $dU=\dbar\mathbbmss{Q}=\dbar \mathbbmss{W}_{\textsc{cd}} = 0$. However, $dS\neq 0$, which means that the whole entropy change is due to the irreversible entropy production. This can be compared with a free expansion process in classical (nonideal) gases. For a further example of a driven qubit, see Ref. \cite{Juan-Delgado21}.

Another example, a quantum damped harmonic oscillator in an environment of oscillators, has been worked out in Ref. \cite{SM}. It appears that at all times the state of the oscillator is a Gibbs state with a time-dependent inverse temperature and a constant Hamiltonian. Because only the eigenvalues of the state change in time and the eigenvectors remain constant, $\dbar\mathbbmss{W}_{\textsc{cd}}$ vanishes. Hence heat and work here reduce to the values obtained from the conventional definitions.

%%%%%%%%%%%%%%%%%%%%%%%%%%%%%%%%%%%%%%%%%%%%%%%%%%%%%%%%%%%%%%%%
\textit{Summary.---}We have revisited the assignment of thermodynamic quantities to an open quantum system strongly coupled to an environment. In general, there is no unique way of separating the system internal energy from that of the environment. Despite this fundamental issue, by introducing a dissipative work, we have shown that it is possible to consistently split the internal energy change into work change (causing no entropy change) and heat change (which can cause entropy change). The key ingredient is to use the trajectory-based description of the state of the system and its associated equation of motion, which is universally valid for any coupling strength and yields a spectral decomposition of the system density matrix, separating the changes in the eigenvalues from those of the eigenvectors. We have compared our entropy-based definitions with the conventional and semiclassical ones and have argued that these two approaches are inadequate. 
%Interestingly, the latter terms correspond to a natural counterdiabatic evolution of the system which forces its state to follow a given trajectory. Further, we have shown that the semiclassical definitions for heat and work, based on expressing the system state in terms of the energy eigenbasis, in general do not properly account for the separation of heat and work. 
More importantly, by using a definition of nonequilibrium temperature in quantum systems, we have obtained the irreversible entropy production in a general dynamical process. We have demonstrated that the irreversible entropy production is a function of the system variables other than the energy and does not have any explicit dependence on the environmental degrees of freedom. We have also derived an inequality which can be contrasted with the generalized Clausius inequality. The consistency of our formalism has been illustrated in paradigmatic scenarios.

%%%%%%%%%%%%%%%%%%%%%%%%%%%%%%%%%%%%%%%%%%%%%%%%%%%%%%%%%%%%%%%%
\textit{Acknowledgement.---}We thank F. J. G\'omez-Ruiz and M. Zahidy for discussions. This work has been supported in part by the Academy of Finland's Center of Excellence QTF Project 312298 and Sharif University of Technology's Office of Vice President for Research and Technology.

\textit{Note added.---}Recently, we learned about another independent related study \cite{Ahmadi}.

%%%%%%%%%%%%%%%%%%%%%%%%%%%%%%%%%%%%%%%%%%%%%%%%%%%%%%%%%%%%%%%%
%\bibliography{e-bib-2}	
%apsrev4-2.bst 2019-01-14 (MD) hand-edited version of apsrev4-1.bst
%Control: key (0)
%Control: author (8) initials jnrlst
%Control: editor formatted (1) identically to author
%Control: production of article title (0) allowed
%Control: page (0) single
%Control: year (1) truncated
%Control: production of eprint (0) enabled
%

%%%%%%%%%%%%%%%%%%%%%%%%%%%%%%%%%%%%%%%%%%%%%%%%%%%%%%%%%%%%%%%%
\newpage
\onecolumngrid
\appendix
\begin{center}
\textbf{Supplemental Material}
\end{center}
%%%%%%%%%%%%%%%%%%%%%%%%%%%%%%%%%%%%%%%%%%%%%%%%%%%%%%%%%%%%%%%%
\section{Irreversible entropy production} 
\label{irr}

Irreversible entropy production is defined as the difference between the entropy $dS$ and the reversible entropy. Using the temperature of the system $T$ as introduced in Ref. \cite{qtemperature} and Eq. (18) of the main text the irreversible entropy production is given by $\dbar \mathbbmss{S} =dS-\beta \,\dbar\mathbbmss{Q}$. By expanding $\dbar \varrho^{(\mathrm{ev})}$ in terms of the set of orthonormal traceless observables $\{O_{i}\}$ (described in the main text) as 
\begin{equation}
\dbar \varrho^{\mathrm{(ev)}}= \textstyle{\sum_{i}} \dbar x_{i} \,O_{i},
\end{equation}
where $\dbar  x_{i}=\mathrm{Tr}[\dbar \varrho^{(\mathrm{ev})} O_{i}]$ we can obtain from the entropy change 
$dS=-\mathrm{Tr}[\dbar \varrho^{\mathrm{(ev)}} \ln \varrho]$ that
\begin{align}
d S=&-\mathrm{Tr}[\dbar\varrho^{\mathrm{(ev)}} O_1] \,\mathrm{Tr}[O_1 \ln\varrho]-\textstyle{\sum_{i\geqslant 2}}\mathrm{Tr}[\dbar \varrho^{\mathrm{(ev)}} O_{i}] \mathrm{Tr}[O_{i}\ln\varrho] \nonumber\\
=&-\frac{1}{\sqrt{\mathrm{Cov}(H,H)}}\mathrm{Tr}[O_1 \ln\varrho]\, \dbar\mathbbmss{Q}-\textstyle{\sum_{i\geqslant 2}}\mathrm{Tr}[\dbar \varrho^{\mathrm{(ev)}} O_{i}] \mathrm{Tr}[O_{i}\ln\varrho], \nonumber\\
=&\beta \,\dbar\mathbbmss{Q}-\textstyle{\sum_{i\geqslant 2}}\mathrm{Tr}[\dbar \varrho^{\mathrm{(ev)}} O_{i}] \mathrm{Tr}[O_{i}\ln\varrho],\nonumber\\
=&\beta \,\dbar\mathbbmss{Q}-\textstyle{\sum_{i\geqslant 2}} \sum_{j} dr_{j} \langle r_{j} |O_{i} |r_{j}\rangle \mathrm{Tr}[O_{i}\ln\varrho],
\end{align}
where we have used the definition of inverse temperature $\beta$ [Eq. (18) of the main text]. The expression for the irreversible entropy production is obtained as 
\begin{align}
\dbar \mathbbmss{S} =&-\textstyle{\sum_{i\geqslant 2}}\mathrm{Tr}[\dbar  \varrho^{\mathrm{(ev)}} O_{i}] \,\mathrm{Tr}[O_{i}\ln\varrho]\nonumber
\\
=&-\textstyle{\sum_{i\geqslant 2}} \sum_{j} dr_{j} \langle r_{j} |O_{i} |r_{j}\rangle \mathrm{Tr}[O_{i}\ln\varrho].
\label{entro_prod-2}
\end{align}

It should be noted that despite what may seem at first, $\dbar\mathbbmss{S}$ is independent of the choice of the basis operators $\{O_{i}\}_{i\geqslant 2}$. This can be inferred from the fact that the entropy change $dS=-\mathrm{Tr}[d\varrho\ln\varrho]$ is independent of the choice of basis, and $\dbar\mathbbmss{Q}$ (heat depends only on $O_{1}$). Thus, one can conclude that the choice of observables $\{O_{i\geqslant 2}\}$ does not affect the irreversible entropy production. We also prove this observation in a different manner in the following. 

To see that $\dbar\mathbbmss{S}$ is independent of our choice of basis operators $\{O_{i}\}_{i=0}^{D^2-1}$, let $\{O'_{i}\}_{i=0}^{D^2-1}$ be another set of orthonormal operators such that still $O'_{0}=O_{0}$ and $O'_{1}=O_{1}$. These bases must be related through an orthogonal matrix $\mathcal{U}$ as $O_{i}=\sum_{k=2}^{D^{2}-1}\mathcal{U}_{ik}O'_{k}$, for $i\geqslant 2$. From Eq. (\ref{entro_prod-2}) one finds
\begin{align*}
\dbar \mathbbmss{S}&= - \textstyle{\sum_{i=2}^{D^2 -1}} \mathrm{Tr}[O_{i}\ln \varrho]\,\mathrm{Tr}[O_{i} \dbar \varrho^{\mathrm{(ev)}}]\\
&=- \textstyle{\sum_{i=2}^{D^2 -1}} \textstyle{\sum_{k=2}^{D^2 -1}} \textstyle{\sum_{l=2}^{D^2 -1}} \mathcal{U}_{ik}\mathcal{U}_{il}\mathrm{Tr}[O'_{k}\ln \varrho]\,\mathrm{Tr}[O'_{l} \delta \varrho^{\mathrm{(ev)}}]\\
&= -  \textstyle{\sum_{k=2}^{D^2 -1}} \textstyle{\sum_{l=2}^{D^2 -1}} \big( \textstyle{\sum_{i=2}^{D^2 -1}} (\mathcal{U}^{T})_{ki}(\mathcal{U})_{il} \big)\mathrm{Tr}[O'_{k}\ln \varrho]\,\mathrm{Tr}[O'_{l} \dbar \varrho^{\mathrm{(ev)}}]\\
&=- \textstyle{\sum_{k=2}^{D^2-1}} \mathrm{Tr}[O'_{k}\ln \varrho]\,\mathrm{Tr}[O'_{k} \dbar \varrho^{\mathrm{(ev)}}],
\end{align*}
which proves $\dbar \mathbbmss{S}$ is independent of the choice of basis. 

\textit{Remark.}---One can observe that \cite{qtemperature} $\beta$ appears naturally as a variable of the state $\varrho$ as
\begin{equation}
\varrho=(1/Z) e^{-\beta H + \sum_{i\geqslant 2}c_{i} O_{i}},
\end{equation}
where $Z=e^{-\mathrm{Tr}[\mathbbmss{H}]}$ and $c_{i}=-\mathrm{Tr}[O_{i}\mathbbmss{H}]$.

%%%%%%%%%%%%%%%%%%%%%%%%%%%%%%%%%%%%%%%%%%%%%%%%%%%%%%%%%%%%%%%%
\section{Further justification for $\dbar \mathbbmss{S}$}

\begin{figure}[tp]
\includegraphics[width=0.22\linewidth]{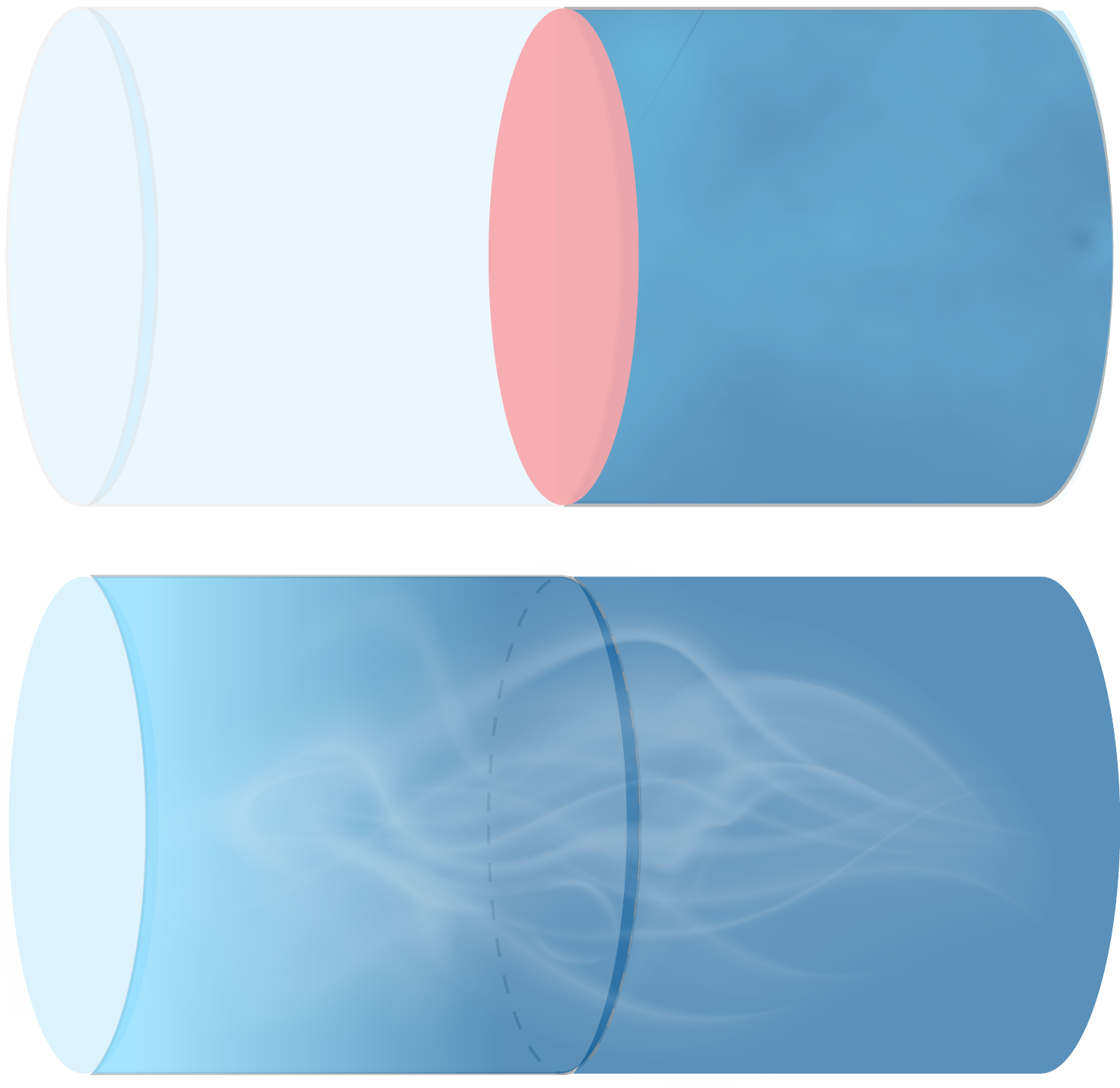}
\hskip4cm
\includegraphics[width=0.18\linewidth]{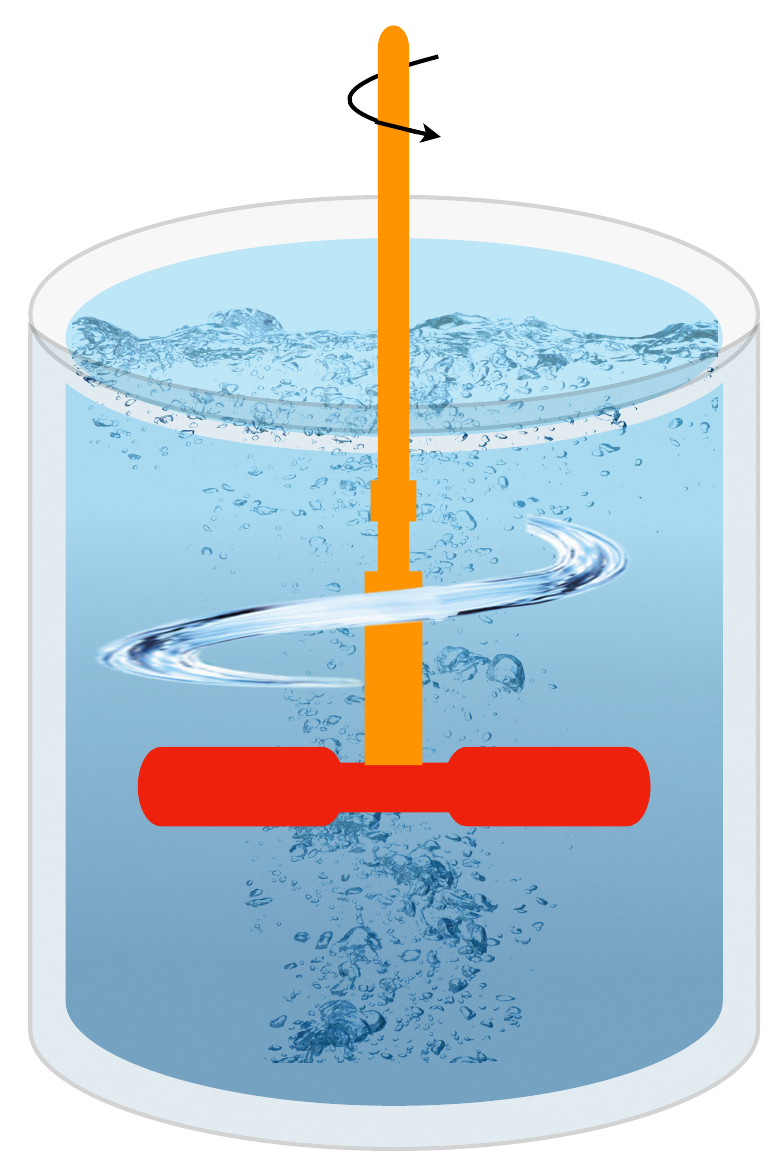}
\caption{(Left) Schematic of the free expansion. An ideal gas is confined in a half of an adiabatic container by using a partition. The partition is then removed and the gas fills the whole container. (Right) Schematic of Joule's experiment. By rotating paddles we perform work on a fluid inside an adiabatic container.}
\label{FJ-exp}
\end{figure}

Here, to argue that $\dbar \mathbbmss{S}$, as in Eq. (19) of the main text, can well describe the irreversible entropy production, we revisit two classic processes in thermodynamics: free expansion and Joule's experiment \cite{book:Finn}---Fig. \ref{FJ-exp}.  

%%%%%%%%%%%%%%%%%%%%%%%%%%%%%%%%%%%%%%%%%%%%%%%%%%%%%%%%%%%%%%%%
\subsection{Free expansion}

Assume an isolated rigid container of an ideal gas such that the gas is initially confined in one half of the container, separated by an almost massless partition from the other half---no energy exchange with the environment. The partition is then removed suddenly and the gas can fill the whole volume of the container. This process does not involve any energy change, since the temperature is constant (both heat and work are zero). However, pressure and volume are two other variables of the system---one of which can be considered as an independent variable since the state equation connects these variables. Although no energy change occurs in the system, because the change of pressure (and volume) causes irreversible entropy production in the system.

%%%%%%%%%%%%%%%%%%%%%%%%%%%%%%%%%%%%%%%%%%%%%%%%%%%%%%%%%%%%%%%%
\subsection{Joule's experiment}

Assume a fluid (a liquid or a nonideal gas) inside a container with adiabatic (i.e., thermally isolated) walls. By rotating paddles from outside of the container it is possible to perform work on the fluid, which leads to an increase in the internal energy. At the same time rotating paddles also leads to a change in other independent variables and degrees of freedom of the system, such as angular momentum, density, pressure, which in turn yield an increase in the disorder in the system and hence an increase in the irreversible entropy production.

%%%%%%%%%%%%%%%%%%%%%%%%%%%%%%%%%%%%%%%%%%%%%%%%%%%%%%%%%%%%%%%%

%%%%%%%%%%%%%%%%%%%%%%%%%%%%%%%%%%%%%%%%%%%%%%%%%%%%%%%%%%%%%%%%
\section{A Clausius inequality for the irreversible entropy production $\dbar \mathbbmss{S}$}

By using the identity $\mathrm{Tr}[X]\,\mathrm{Tr}[Y]= \mathrm{Tr}[X \otimes Y]$, for any pair of linear operators $X$ and $Y$, we can rewrite Eq. (19) of the main text as
\begin{equation}
\dbar\mathbbmss{S} =\mathrm{Tr}[(\mathbbmss{H}\otimes \dbar\varrho^{(\mathrm{ev})})\mathbbmss{O}],
\end{equation}
where $\mathbbmss{O} = \sum_{i \geqslant 2} O_{i}\otimes O_{i}$. Note that $\mathbbmss{H}$ is a positive operator and  
%and $\dbar \varrho^{(\mathrm{ev})}=\sum_{k}dr_{k}|r_{k}\rangle \langle r_{k}|=\sum_{k|dr_{k}\geqslant 0} dr^{(+)}_{k}|r_{k}\rangle \langle r_{k}| - \sum_{k|dr_{k}\leqslant 0} dr^{(-)}_{k}|r_{k}\rangle \langle r_{k}| $, where $dr^{(\pm)}_{k}\geqslant 0$. 
$\mathbbmss{H}\otimes \dbar\varrho^{(\mathrm{ev})}$ and both $\mathbbmss{O}$ 
%=\sum_{l} \mathbbmss{o}_{l}|\mathbbmss{o}_{l} \rangle \langle \mathbbmss{o}_{l}|$ 
are traceless Hermitian operators. We also use the fact that any Hermitian operator $X$ can be written as $X=X^{(+)}-X^{(-)}$ where both $X^{(\pm)}$ are positive operators, defined in the spectral form as $X^{(+)} = \sum_{k|x_{k\geqslant 0}} x_{k}|x_{k} \rangle \langle x_{k}|$ and $X^{(-)} = -\sum_{k|x_{k\leqslant 0}} x_{k}|x_{k} \rangle \langle x_{k}|$. If such an operator is also traceless, then we have $\mathrm{Tr}[X^{(+)}] = \mathrm{Tr}[X^{(-)}]$. As a result, it follows that 
\begin{align}
\dbar\mathbbmss{S} &=\mathrm{Tr}\big[\big\{(\mathbbmss{H}\otimes \dbar\varrho^{(\mathrm{ev})})^{(+)} - (\mathbbmss{H}\otimes \dbar\varrho^{(\mathrm{ev})})^{(-)} \big\} \mathbbmss{O} \big] \nonumber\\
&= \textstyle{\sum_{l}} \mathbbmss{o}_{l} \langle \mathbbmss{o}_{l}|(\mathbbmss{H}\otimes \dbar\varrho^{(\mathrm{ev})})^{(+)}|\mathbbmss{o}_{l} \rangle - \textstyle{\sum_{l}} \mathbbmss{o}_{l} \langle \mathbbmss{o}_{l}|(\mathbbmss{H}\otimes \dbar\varrho^{(\mathrm{ev})})^{(-)}|\mathbbmss{o}_{l} \rangle \nonumber\\
& \geqslant \mathbbmss{o}_{\min}\, \textstyle{\sum_{l}} \langle \mathbbmss{o}_{l}|(\mathbbmss{H}\otimes \dbar\varrho^{(\mathrm{ev})})^{(+)}|\mathbbmss{o}_{l} \rangle - \mathbbmss{o}_{\max} \,\textstyle{\sum_{l}} \langle \mathbbmss{o}_{l}|(\mathbbmss{H}\otimes \dbar\varrho^{(\mathrm{ev})})^{(-)}|\mathbbmss{o}_{l} \rangle \nonumber\\
& =  -(\mathbbmss{o}_{\max} - \mathbbmss{o}_{\min}) \mathrm{Tr}[(\mathbbmss{H}\otimes \dbar\varrho^{(\mathrm{ev})})^{(+)}].\nonumber
\end{align}
Thus,
\begin{equation}
\dbar \mathbbmss{S} + b(\mathbbmss{O}) \,\dbar B\geqslant 0,
\label{Cl-ineq}
\end{equation}
where $b(\mathbbmss{O}) = |\mathbbmss{o}_{\max} - \mathbbmss{o}_{\min}|$ is spread of the operator $\mathbbmss{O}$, which somehow encompasses the information of all non-Hamiltonian system observables needed to describe the state of the system. This quantity is independent of $\varrho$. In addition, $\dbar B = \mathrm{Tr}[(\mathbbmss{H}\otimes \dbar\varrho^{(\mathrm{ev})})^{(+)}] = - (\sum_{k}\ln r_{k})(\sum_{k'}dr^{(+)}_{k'})\geqslant 0$, which fully depends on the state $\varrho$.

The above general inequality reminisces a \textit{generalized Clausius inequality}---see Ref. [62] of the main text. We remark that all quantities appeared here are explicitly system or trajectory dependent, without need for explicit parameters from an environment or heat bath. Of course when the trajectory is not unitary, existence of an environment is inevitable, hence $\dbar B$ can be considered to have contributions from the environment.

%%%%%%%%%%%%%%%%%%%%%%%%%%%%%%%%%%%%%%%%%%%%%%%%%%%%%%%%%%%%%%%%
\section{Alternative form for the irreversible entropy production}

Consider the relative entropy
\begin{align}
S(\varrho\|\varrho_{\mathrm{eq}})=\mathrm{Tr}[\varrho\ln \varrho-\varrho\ln \varrho_{\mathrm{eq}}],
\end{align}
where $\varrho_{\mathrm{eq}}$ is a reference equilibrium state which is a Gibbs state with the temperature $\beta$ of the system in the form $\varrho_{\mathrm{eq}}=e^{-\beta H}/Z_{\mathrm{eq}}$, where $H$ is the  physical Hamiltonian of the system and $Z_{\mathrm{eq}}=\mathrm{Tr}[e^{-\beta H}]$. Using $\varrho=\sum_{k} r_{k}|r_{k} \rangle\langle r_{k} |$, it follows that
\begin{align}
S(\varrho\|\varrho_{\mathrm{eq}}) =& \textstyle{\sum_{k}} r_{k} \big(\ln r_{k}-\langle r_{k}|\ln \varrho_{\mathrm{eq}}|r_{k} \rangle\big) \nonumber\\
=& \textstyle{\sum_{k}} r_{k} \ln r_{k} + \beta \mathrm{Tr}[\varrho H]+\ln Z_{\mathrm{eq}}.
\end{align}
Differentiation with respect to time yields
\begin{align}
dS(\varrho\|\varrho_{\mathrm{eq}})=&\textstyle{\sum_{k}} dr_{k} \big(\ln r_{k}-\langle r_{k}|\ln \varrho_{\mathrm{eq}}| r_{k} \rangle\big) -\sum_{k} r_{k} \big(\langle dr_{k}|\ln \varrho_{\mathrm{eq}}|r_{k} \rangle -\beta\langle r_{k}|dH|r_{k}\rangle - d\beta \langle r_{k}|H|r_{k}\rangle - dZ_{\mathrm{eq}}/Z_{\mathrm{eq}} + \langle r_{k}|\ln \varrho_{\mathrm{eq}}| dr_{k} \rangle\big).
\end{align}

Using the relations
\begin{gather}
dS=- \textstyle{\sum_{k}} dr_{k} \ln r_{k},\\
\dbar \mathbbmss{W}_{\textsc{cd}}= \textstyle{\sum_{k}} r_{k} \left(\langle dr_{k}|H|r_{k} \rangle+\langle r_{k}|H|dr_{k} \rangle\right),\\
\dbar \mathbbmss{Q} =\textstyle{\sum_{k}} dr_{k} \langle  r_{k} |H| r_{k}  \rangle, \\
dZ_{\mathrm{eq}}/Z_{\mathrm{eq}} = -d\beta\,\mathrm{Tr}[H \varrho_{\mathrm{eq}}] - \beta\,\mathrm{Tr}[dH\varrho_{\mathrm{eq}}],
\end{gather}
it follows that
\begin{align}
\dbar \mathbbmss{S} =-dS(\varrho\|\varrho_{\mathrm{eq}})+\beta\, \dbar\mathbbmss{W}_{\textsc{cd}} +\beta\,\mathrm{Tr}[(\varrho-\varrho_{\mathrm{eq}})\,dH] + d\beta\,\mathrm{Tr}[(\varrho-\varrho_{\mathrm{eq}}) H],
\end{align}
where we have used the definition of irreversible entropy change $\dbar \mathbbmss{S} = dS-\beta \, \dbar \mathbbmss{Q}$.

However, under general conditions of a Markovian dynamics with a time-dependent steady state of $\varrho_{\mathrm{eq}}$ where both $\beta$ and $H$ can be time-dependent, from $dS(\varrho\|\varrho_{\mathrm{eq}})\leqslant 0$ it can be concluded that 
\begin{align}
\dbar \mathbbmss{S} \geqslant \beta\, \dbar \mathbbmss{W}_{\textsc{cd}} -\beta\,\mathrm{Tr}[(\varrho-\varrho_{\mathrm{eq}})\,dH]-d\beta\,\mathrm{Tr}[(\varrho-\varrho_{\mathrm{eq}}) H].
\end{align}

%%%%%%%%%%%%%%%%%%%%%%%%%%%%%%%%%%%%%%%%%%%%%%%%%%%%%%%%%%%%%%%%
\section{Temperature in the irreversible entropy production and the Clausius inequality}
\label{sec:temp}

The entropy production itself originates from the Clausius inequality. A widespread assumption in thermodynamics is that the temperature $T=1/\beta$ in the Clausius inequality $\oint \dbar Q/T \leqslant 0$, and hence the one in the irreversible entropy production ($d\mathbbmss{S}=dS-\beta \dbar Q$), is the temperature of the ambient bath. However, it can be argued that in general scenarios this assignment may be misleading. In fact, the discussion about the meaning of $T$ in $\dbar Q/T$ even in standard classical thermodynamics is far from settled. For example, see the discussion in Ref. \cite{Evans}. By revisiting the proof of the Clausius inequality (see, e.g., Ref. \cite{book:Finn}) it becomes clear that the temperature that appears in this inequality is the temperature of an \textit{auxiliary} environment which can provide the system with $\dbar Q$ (the heat needed for proceeding with the given thermodynamics process). Hence there is some freedom in this temperature as far as the auxiliary environment can provide the necessary heat for the process. Even if there is a real physical environment around the system, there is no need to put the temperature of this environment into the entropy production to have a valid Clausius inequality. 

Given this freedom, the temperature in the Clausius inequality can instead be assigned to the system itself as its instantaneous (and generally nonequilibrium) temperature. Clausius himself had indicated in his 1856 paper: ``$T$ is a function of the temperature of the changing body in that moment in which it takes in the element of heat $\dbar Q$ or, if the body has different temperatures in its different parts, of the temperature of that part of it which takes in $dQ$'' \cite{pap} (see also Ref. \cite{Clausius}). Along similar lines we also think that the correct way to describe the entropy production is when the temperature therein is the system temperature. This is in accordance with a system-based description of the thermodynamic properties and is in complete accordance with the laws of thermodynamics. As we have shown in the main text and also in Sec. \ref{irr} (of this Supplemental Material) the system temperature---in the way that we have derived---appears naturally when expanding entropy in terms of a set of relevant independent variables. Thus, with our definition, entropy production has a reasonable and natural physical interpretation as the contribution of the other properties and physical variables of the system but heat (or energy) in entropy change of the system. In fact, if in general scenarios we put the temperature of the environment into the definition of the entropy production, the term $\dbar Q/T$ would not be a genuine part of the entropy change of the \textit{system}. It seems more reasonable that entropy production should be a \textit{system} property not depending on hypothetical situations by which a particular process can be realized.

%%%%%%%%%%%%%%%%%%%%%%%%%%%%%%%%%%%%%%%%%%%%%%%%%%%%%%%%%%%%%%%%
\section{Details of Example I}

%%%%%%%%%%%%%%%%%%%%%%%%%%%%%%%%%%%%%%%%%%%%%%%%%%%%%%%%%%%%%%%%
\subsection{Case (i)}

The initial state of the atom is a thermal state $\varrho(0)=e^{-\beta(0)\omega_{0}\sigma_{z}}/\mathrm{Tr}[e^{-\beta(0)\omega_{0}\sigma_{z}}]$ with initial inverse temperature $\beta(0)$. In this case, the trajectory becomes 
\begin{equation}
\varrho(t)=(1/2) \textstyle{\sum_{k=0,1}}\left( 1+(-1)^{k+1}e^{-2 \gamma t} \tanh(\beta(0) \omega_{0})\right) |r_{k}\rangle\langle r_{k}|,
\end{equation}
where $\{|r_{0}\rangle=|0\rangle,|r_{1}\rangle=|1\rangle\}$. Since $\varrho(t)$ is diagonal in a time-independent basis, its change is only due to the change in the eigenvalues and thus the whole internal energy change is due to the heat exchange with the bath, $\dot{\mathbbmss{Q}}(t)=\dot{U}(t)= 2\gamma \omega_{0} e^{-2\gamma t} \tanh [\beta(0) \omega_{0}]$. A direct calculation also shows that $\dot{q}(t)=\dot{\mathbbmss{Q}}(t)$ and $\dot{w}(t)=\dot{\mathbbmss{W}}(t)$. 

%%%%%%%%%%%%%%%%%%%%%%%%%%%%%%%%%%%%%%%%%%%%%%%%%%%%%%%%%%%%%%%%
\subsection{Case (ii)}

The initial state of the atom is a pure state $\varrho(0)=(1/2)(\openone+\sigma_{x})$, which is a steady state for the dissipative part of the dynamics $\mathpzc{D}[\varrho(0)]=0$. However, the time evolution of the state, i.e., $\varrho(t)$ is not an steady state of the dissipative part and is given by 
\begin{align}
 \varrho(t)=(1/2)\openone + c(t)|0\rangle\langle 1|+c^{\ast}(t)|1 \rangle\langle 0|,
\end{align}
where 
\begin{equation}
c(t)=e^{-\gamma t} \big(\Delta  \cosh (t\Delta)+(\gamma -2i\omega_{0}) \sinh (t\Delta) \big)/ (2 \Delta),
\end{equation}   
and $\Delta=\sqrt{\gamma^2-4 \omega_{0}^2}$. From the expression for $\dot{U}(t)$ in the main text,
\begin{equation}
\dot{U}(t)=-2\gamma \omega_{0} \mathrm{Tr}[\sigma_{z}\varrho(t)],
\end{equation}
it is evident that although the state of the system varies in time, the system internal energy does not change, $\dot{U}(t)=0$.
   
By calculating the eigenvalues and the eigenvectors of $\varrho(t)$ as follows:
\begin{align}
&r_{\pm}(t)=\frac{1}{2} \pm \frac{e^{-\gamma t}}{2\Delta }\left(\gamma ^2 \cosh (2 t\Delta)+\gamma 
\Delta  \sinh (2 t\Delta)-4 \omega _{0}^2\right)^{1/2},\\
&\sqrt{N_{\pm}(t)} |r_{\pm}(t)\rangle = \frac{\big(\gamma^{2} \cosh (2 t\Delta)+\gamma  \Delta  \sinh (2t \Delta) - 4
\omega_{0}^2 \big)^{1/2}}{\Delta  \cosh (t\Delta)+(\gamma +2 i \omega_{0}) \sinh(t\Delta)} |0\rangle  \pm |1\rangle,
\end{align}
where $N_{\pm}$ is the normalization factor for $|r_{\pm}(t)\rangle$, it can be shown that $\dot{\mathbbmss{Q}}(t)=\dot{\mathbbmss{W}}(t)=\dot{q}(t)=\dot{w}(t)=0$. 

%%%%%%%%%%%%%%%%%%%%%%%%%%%%%%%%%%%%%%%%%%%%%%%%%%%%%%%%%%%%%%%%
\subsection{Case (iii)}

Starting from $\varrho(0)=(\openone+[\sigma_{x}+\sigma_{z}]/2)/2$ and obtaining the instantaneous state of the system, we can obtain the heat, work, and internal energy change in the system. Figure 1 of the main text illustrates the results. It can be shown that $\dot{q}(t)=\dot{U}(t)=-(\gamma/2) e^{-2 \gamma t}$ and $\dot{w}(t)=0$. 

%%%%%%%%%%%%%%%%%%%%%%%%%%%%%%%%%%%%%%%%%%%%%%%%%%%%%%%%%%%%%%%%

%%%%%%%%%%%%%%%%%%%%%%%%%%%%%%%%%%%%%%%%%%%%%%%%%%%%%%%%%%%%%%%%
\begin{figure}[tp]
\includegraphics[width=0.47\linewidth]{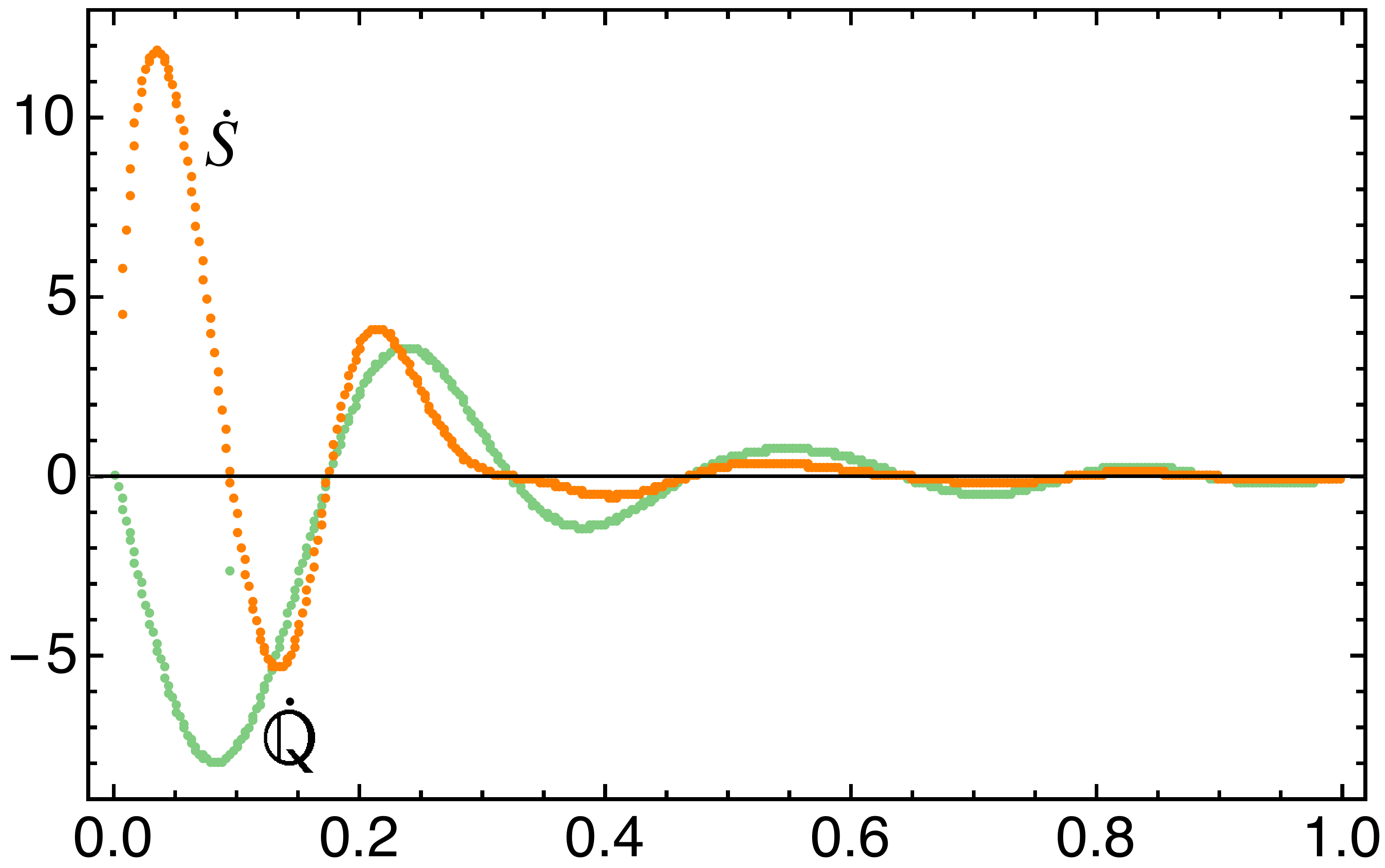}
\caption{Numerical simulation for TB-STA heat exchange rate (green)  and entropy rate (orange) vs. time, for a harmonic oscillator in a bath of harmonic oscillators as given in Example III, for $M=255$, $g_{k}=\sqrt{\omega_{k}/\pi}e^{-\omega_{k}/(2\omega_c)}$, $\omega_{k}=0.1 k$, $\omega_{0}=1$, and $\omega_c=5$.} 
\label{figure3}
\end{figure}
%%%%%%%%%%%%%%%%%%%%%%%%%%%%%%%%%%%%%%%%%%%%%%%%%%%%%%%%%%%%%%%%

%%%%%%%%%%%%%%%%%%%%%%%%%%%%%%%%%%%%%%%%%%%%%%%%%%%%%%%%%%%%%%%%
\section{Example III: Damped harmonic oscillator within a bath of oscillators}
\label{example-oscillator}

The main manuscript discusses Examples I and II. We next consider a quantum harmonic oscillator interacting with a bath of oscillators with the total Hamiltonian 
\begin{align}
H_{\mathsf{SB}}=\omega_{0} \hat{a}^{\dag} \hat{a}+ \textstyle{\sum_{k=1}^{M}}\omega_{k} \hat{b}_{k}^{\dag} \hat{b}_{k}+ \textstyle{\sum_{k=1}^{M}} g_{k} (\hat{a}^{\dag} \hat{b}_{k}+ \hat{a} \hat{b}_{k}^{\dag}),
\end{align}
where $M$ is the number of the bath oscillators. For simplicity of the analysis, we assume the initial system-bath state $ |1\rangle_{\mathsf{S}}\otimes |0\rangle^{\otimes M}_{\mathsf{B}}$, where $|i\rangle$ denotes the eigenstate of the corresponding number operator with eigenvalue $i$. Using the exact numerical simulation, we obtain the time derivatives of heat and entropy; see Fig. \ref{figure3}. Interestingly, we see that in this special case, $\dbar \mathbbmss{S}$ vanishes. Hence the process is thermodynamically reversible. A closer look at the dynamics shows that the system evolves quasistatically and its state at any time is given by a Gibbs state $\varrho_{\mathsf{S}}(t)=\varrho_{\mathrm{eq}}(t) = e^{-\beta(t) H_{\mathsf{S}}}/\mathrm{Tr}[e^{-\beta(t) H_{\mathsf{S}}}]$ and $\beta(t)$ is obtained from Eq. (18) of the main text. Thus, only the eigenvalues of the state change in time and the eigenvectors remain unchanged, indicating that $\dbar \mathbbmss{W}_{\textsc{cd}}$ vanishes. Hence heat and work here are consistent  with the conventional definitions.

%%%%%%%%%%%%%%%%%%%%%%%%%%%%%%%%%%%%%%%%%%%%%%%%%%%%%%%%%%%%%%%%
%
%%%%%%%%%%%%%%%%%%%%%%%%%%%%%%%%%%%%%%%%%%%%%%%%%%%%%%%%%%%%%%%%

\end{document}